\documentclass[aps,prb,twocolumn,groupedaddress]{revtex4}
\usepackage{amssymb}
\usepackage{graphicx}
\usepackage{epsfig}
\usepackage{float}
\usepackage{amsmath}

\begin{document}

\title{\bf A critical assessment of the pairing symmetry in  Na$_x$CoO$_2\cdot y$H$_2$O}
\author{I.I. Mazin, M.D. Johannes}

\affiliation{Code 6391, Naval Research Laboratory, Washington, D.C. 20375} 

\begin{abstract}
We examine each of the symmetry-allowed pairing states of Na$_x$CoO$_2\cdot y$H$_2$O and compare their 
properties to what is experimentally and theoretically established about the compound.  In this way, we 
can eliminate the vast majority of states that are technically allowed and narrow the field to two, both 
of $f$-wave type states.  We discuss the expected features of these states and suggest experiments that 
can distinguish between them.  We also discuss odd-frequency gap pairing and how it relates to available 
experimental evidence
\end{abstract}

\maketitle

Na$_x$CoO$_2\cdot y$H$_2$O is a novel superconductor which, despite a relatively low superconducting temperature
\cite{KTHS+03} of only $\sim$ 5K, has recently attracted substantial experimental and theoretical attention.  Much of the
interest is driven by an as-of-yet unresolved pairing state that is presumed to be highly unusual and possibly (as we will
argue in this Letter, likely) even more unconventional than the $d$-wave superconductivity of the cuprates or $p$-wave
superconductivity of the ruthenates.  A survey of the current literature for Na$_x$CoO$_2\cdot y$H$_2$O reveals that in the
two years since its discovery, various groups have proposed $s$-wave\cite{YKMY,MYHW+03,KKYT03}, $p$-wave $\mathbf{z}(x+iy)$
\cite{MKCM+04,YYMM05}, $d$-wave $x^{2}-y^{2}+2ixy$\cite {DSMS04,OIM03,TWHY+04,YYMM05} and various versions of $f$ \cite
{KKYT03,YYMM05,WHKO+03} and $i$-wave states, as well as an odd-frequency triplet $s$ state \cite{MDJ+b}.  To our knowledge,
Na$_{x}$CoO$_{2}\cdot y$H$_{2}$O holds the record for the greatest number of different superconducting symmetries proposed for
one compound.

Synthesis of single crystal Na$_x$CoO$_2\cdot y$H$_2$O is difficult and polycrystalline samples often exhibit inhomogeneities
in Na distribution and H$_2$O accumulation \cite{DPC+04,BGU+04}.  The compound is furthermore chemically unstable at ambient
temperature and humidity \cite{MLF+03}, making it difficult to handle and characterize.  For these reasons, well-reproducible
and reliable experimental results that could be expected to unravel the precise superconducting state have been slow to
emerge.  Still, there are several experimental facts that are rather well established, reproducible and which bear immediate
relevance to the superconducting symmetry.  Other facts that follow from confirmed knowledge about the crystal and electronic
structure allow the exclusion of at least a few of the symmetry-allowed pairing states.  An absence of knowledge about which
states conform to symmetry requirements, which are excluded by experiment and which are physically unreasonable has
frequently lead both theorists and experimentalists to concentrate on those pairing symmetries which are compatible with a
specific data set or a specific theory of pairing to the detriment of a broader, consistent picture.

In this Letter, we list all the different symmetry representations that are compatible with a hexagonal crystal
structure, according to the seminal work of Sigrist and Ueda\cite{MSKU91}.  Based on what is currently known
experimentally about Na$_x$CoO$_2\cdot y$H$_2$O, we discuss which states can be eliminated from
consideration with a reasonable degree of confidence.  We will show that surprisingly few candidates survive
this scrutiny, and that all of these are highly unconventional and, in a sense, more exotic than either the
high-T$_c$ cuprates or the $p$-wave ruthenates.  The list of allowed symmetries \cite{MSKU91} up to L=3
({\it i.e.} up to the $f$-states) is given in Table \ref{sym}.

\begin{table}
\caption{Symmetry-allowed pairing states for hexagonal symmetry}  
\begin{tabular*}{0.90 \linewidth}{@{\extracolsep{\fill}}llcccc}
&  & 2D & DOS & $\mu $SR &  \\
1 & 1 & Y & N & Y & $s$ \\
2 & $x^{2}+y^{2}$ & Y & N & Y & $s$ \\
3 & $z^{2}$ & N & Y & Y & $s$ \\
4 & $x\mathbf{\hat{z}}^{(a)}$ & Y & Y & Y & $p$ \\
5 & $y\mathbf{\hat{z}}^{(a)}$ & Y & Y & Y & $p$ \\
6 & ($x\pm iy)\mathbf{\hat{z}}$ & Y & N & N & $p$ \\
7 & $z\mathbf{\hat{x}}$ & N & Y & Y & $p$ \\
8 & $z\mathbf{\hat{y}}$ & N & Y & Y & $p$ \\
9 & $z(\mathbf{\hat{x}\pm }i\mathbf{\hat{y}})$ & N & N & N & $p$ \\
10 & $y\mathbf{\hat{x}+}x\mathbf{\hat{y}}^{(b)}$ & Y & N & Y & $p$ \\
11 & $x\mathbf{\hat{x}-}y\mathbf{\hat{y}}$ & Y & N & Y & $p$ \\
12 & ($\mathbf{\hat{x}\pm }i\mathbf{\hat{y})}(x\pm iy\mathbf{)}$ & Y & Y & N
& $p$ \\
13 & $x\mathbf{\hat{x}+}y\mathbf{\hat{y}}$ & Y & N & Y & $p$ \\
14 & $y\mathbf{\hat{x}-}x\mathbf{\hat{y}}^{(c)}$ & Y & N & Y & $p$ \\
15 & $z\mathbf{\hat{z}}$ & N & Y & Y & $p$ \\
16 & $x^{2}-y^{2(a)}$ & Y & Y & Y & $d$ \\
17 & $xy^{(a)}$ & Y & Y & Y & $d$ \\
18 & ($x\pm iy)^{2}$ & Y & N & N & $d$ \\
19 & $xz$ & N & Y & Y & $d$ \\
20 & $yz$ & N & Y & Y & $d$ \\
21 & ($x\pm iy)z$ & N & Y & N & $d$ \\
22 & $x(x^{2}-3y^{2})\mathbf{\hat{z}}$ & Y & Y & Y & $f$ \\
23 & $y(y^{2}-3x^{2})\mathbf{\hat{z}}$ & Y & Y & Y & $f$ \\
24 & $z[(x^{2}-y^{2})\mathbf{\hat{x}}+2xy\mathbf{\hat{y}}]$ & N & Y & Y & $f$
\\
25 & $z[(x^{2}-y^{2})\mathbf{\hat{y}}+2xy\mathbf{\hat{x}}]$ & N & Y & Y & $f$
\\
&  &  &  &  &  \\
26 & $s\mathbf{\hat{x}},$ $s\mathbf{\hat{y}}$, or $s\mathbf{\hat{z}}$ & Y & Y
& Y & $s$ \\
&  &  &  &  &  \\
&  &  &  &  &
\end{tabular*}
\flushleft{$^{(a)}$ These states are excluded because of their 
proximity to a fully-gapped state
$^{(b)}$ In Table III of Ref. \onlinecite{MSKU91}, this state is
printed with a typo, which is corrected here.
$^{(c)}$ In Table VI of Ref. \onlinecite{MSKU91}, this state is
printed with a typo, which is corrected here.
}
\label{sym}
\end{table}

There are 25 states in this table, excluding the last one which will be discussed separately later.  
We will show that all but two of them are incompatible with the experimental data.

First, we decide which facts are to be considered as firmly established.  Some potentially
very important probes, such as the temperature dependence of the Knight shift, are still
controversial in the sense that different authors report contradictory results
\cite{YKMY,WHKO+03,kobayashi,MKCM+04}.  We have therefore singled out three pieces of evidence on
which all or practically all publications agree.  These are:

 \textit{Two-dimensionality}. Electronic structure calculations for the hydrated compound show an anisotropy in the Fermi
velocity of at least an order of magnitude \cite{MDJ04}, which is supported by an experimentally measured resistive anisotropy
\cite{RJBCS+03,FCC+03} of 10$^3$ - 10$^4$, corresponding to a Fermi velocity anisotropy of 30 to 100 (the resistivity
anisotropy of the unhydrated, high Na content compound, Na$_{0.75}$CoO$_{2},$ which should be substantially lower than that of
Na$_{0.3}$CoO$_{2}\cdot y$H$_{2}$O, was found\cite{BCS+04} to be as high as 500), indicating that the transport along $c$ is
probably incoherent. This is firm evidence that the electronic structure is very strongly 2D.

As found experimentally\cite{SPB+04,LMH+04} and explained theoretically \cite{MDJ05}, the magnetic anisotropy of the
unhydrated high-Na compound is very small, primarily because each Co couples with 7 Co atoms in neighboring layers. While
there are no data on the magnetic coupling at $x=0.3,$ nor for the hydrated compound, one can estimate the reduction in
magnetic coupling from the ratio of the squared Fermi velocities, which is about 20. Thus in the hydrated compound, magnetic
interaction should also be 2D.

Finally, in an interesting difference from both the cuprates and ruthenates, the
Co and O phonons should also be 2D in this system. Of course, water
vibration need not be such, but the absence of the hydrogen isotope effect \cite
{RJBCS+03} clearly indicates their irrelevance for the superconducting pairing.

\textit{Therefore, we conclude that the superconducting order parameter in Na }$_{0.3}$\textit{CoO}$_{2}\cdot
y$\textit{H}$_{2}$\textit{O should be 2D.}

 \textit{Absence of superconductivity-induced spontaneous magnetic moments below
}$T_{c}$\textit{. }Some of the superconducting states listed in Table \ref{sym}
(\#9,12) are nonunitary and have a spontaneous magnetization in the superconducting
state. Others (\#6,18,21) break the time-reversal symmetry for a Cooper pair by virtue
of a nonzero pair orbital moment. In both cases, the resulting nonzero local magnetic
moments are supposed to be detectable\cite{MSKU91}. Note that net magnetization is not
present in the latter case, due to domain formation and internal Meissner screening,
but crystallographic defects and grain or domain boundaries should still host 
nonzero local moments. One of the main arguments in favor of the axial
$(x+iy)\mathbf{\hat{z}}$ state in Sr$_{2}$RuO$_{4}$\cite{APM03} was the fact that muon
spectroscopy revealed the appearance of disordered static magnetic moments below
$T_{c}.$ The accepted interpretation of this finding is that the pairing symmetry has a
nonzero orbital moment. Muon spin rotation experiments for
Na$_{0.3}$\textit{CoO}$_{2}\cdot y$\textit{H}$_{2}$ have been reported
\cite{WHKO+03,YJU+04} and no indications of static moments below $T_{c}$ have been
found. To our knowledge, there are no other works reporting detection or non-detection
of static magnetic moments in this compound. We nevertheless feel confident to include
this fact in our compendium for two reasons: i) one of the reports comes from a group
\cite{WHKO+03} which has performed similar measurements on other superconductors and
was previously able to detect local moments in PrOs$_{4}$Sb$_{12}$\cite{YAAT+03}, and
ii) this is a rare example of an experiment in which  poor sample quality makes the
effect more pronounced rather than obscuring it.

\textit{Therefore, we conclude that neither nonunitary nor }$L\neq 0$
\textit{states are possible in Na}$_{0.3}$\textit{CoO}$_{2}\cdot y$\textit{H}
$_{2}$\textit{O }

 \textit{Absence of a finite superconducting gap. }Several experimental groups have reported experiments indirectly
probing the density of states (DOS) in the superconducting phase. Such experiments, primarily calorimetry, were
instrumental in clarifying the symmetry of pairing in such novel superconductors as SrRuO$_{4}$\cite {APM03} and
MgB$_{2}$\cite{MgB2rev}.  These experiments measure the temperature dependence of either specific
heat\cite{HDY+05,NORAF+} or relaxation rates (NMR\cite{YKMY}, NQR\cite{KIYI+,TFGZ+} or $\mu $SR\cite
{WHKO+03,AKAK+03}). In all work that we are aware of, the authors agree that the low temperature behavior of the DOS
is not exponential\cite{note}. As of yet, no group has reported measurements at temperatures low enough to allow for a
reasonably confident determination of the exact temperature dependence, but most authors suggest a $T^{3}$ (line
nodes) behavior for $T\gtrsim 2$ K and finite-DOS linear behavior at lower $T$.

\textit{These results exclude states with a fully developed sizeable gap on
all Fermi surfaces.}

Armed with these three facts, let us now test the 25 states listed above against them.
First, we eliminate all states that have, by symmetry, strong $ z$-dependence of the
order parameter. These are the ten states \#3, 7, 8, 9, 15, 19, 20, 21, 24, 25. Of the
remaining 15 states, \#1, 2, 10, 11, 12, 13, 14, and 18 have no symmetry restriction
that would require them to have node lines or points. In general, there exists the
possibility of accidental rather than symmetry-induced nodes, or regions with a finite
but extremely small order parameter, similar to the so-called \textquotedblleft
extended $s$ \textquotedblright\ state, earlier considered for superconducting
cuprates. In the case of Na$_{0.3}$CoO$_{2}\cdot y$H$_{2}$O$,$, such accidental nodes
seem highly unlikely due to its specific fermiology.  The Fermi surface of
Na$_{0.3}$CoO$_{2}\cdot y$H$_{2}$O consists of one relatively small nearly circular
cylinder, and, possibly (predicted by the theory, but not yet confirmed experimentally)
six tiny pockets surrounding the first Fermi surface. Whether the latter actually exist
or not, a pairing interaction that would be consistent with the hexagonal symmetry and
at the same time enforce sign change of the order parameter on these small Fermi
surfaces should itself change sign with a variation of the wave vector on the order of
0.2--0.25 of the Brillouin zone dimension. It is hardly possible to imagine a
physically meaningful pairing interaction of this sort.

Finally, the $\mu SR$ experiment allows us to exclude the non-unitary state \#12. Note
that several states that we have already excluded because they are allowed
by symmetry to have a full gap, are additionally excluded as having nonzero orbital
moment. This leaves us with the six states: \#4,5, 16, 17, 22 and 23. 
There is good
reason to believe that the first four are \textit{not} realized. As an example, we
consider the first pair, $x\mathbf{\hat{z}}$ and $y\mathbf{ \hat{z}}.$ In the linear
approximation (that is, expanding the free energy to second order in the order
parameter), they are degenerate with each other and with state \#6,
($x+iy)\mathbf{\hat{z}}$ (see Ref. \cite{MSKU91}). Strong coupling effects, the
spin-orbit interaction and other effects can, formally, tilt the energy balance in
favor of these states, but only in the 4th order in the order parameter (the
distinction between the states themselves appears only in the 6th order). However, all
three states belong to the same symmetry representation and the first two have node
lines, while ($x+iy)\mathbf{ \hat{z}}$ has full gap. This should lead to a considerable difference in
pairing energy that favors the fully gapped state for the same amplitude of the order
parameter and makes stabilization of the other states highly questionable.

\begin{figure} \includegraphics[width = .95\linewidth]{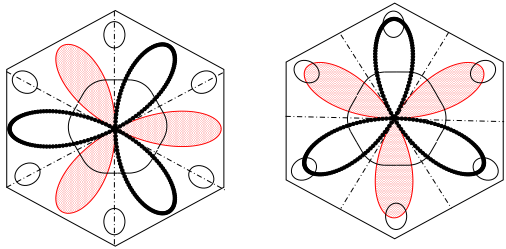} \caption{A schematic of the Na$_{0.3}$CoO$_{2}\cdot
y$H$_{2}$O Fermi surfaces along with the two $f$-wave pairing states that remain after eliminating any states that are
in conflict with experimental evidence.  {\it left panel} $y(y^2-3x^2)\mathbf{\hat{z}}$ has node lines which directly
intersect the theoretically predicted e$_g'$ hole pockets. {\it right panel} The node lines of
$x(x^2-3y^2)\mathbf{\hat{z}}$ fall between the e$_g'$ pockets and therefore cause no loss of pairing energy.} 
\label{fstates} \end{figure}

Excluding states \#4,5,16, and 17 on the basis of this argument, we are
left with \textit{only two possible states: }$x(x^{2}-3y^{2})\mathbf{\hat{z}}
$ and $y(y^{2}-3x^{2})\mathbf{\hat{z}.}$ Assuming only one Fermi surface,
the $a_{1g}$ one around the $\Gamma $ point, we cannot make a further
distinction between the two. However, if one accepts the Fermi surface from
band structure calculations, the latter state has a node line on all $
e_{g}^{\prime }$ Fermi surfaces (See Fig. \ref{fstates}).  Given the small size of these pockets,
the line nodes cause a near loss of pairing for 2/3 of all electrons at the Fermi level,
which is energetically unlikely. Therefore, we conclude that if the $e'_g$ derived
Fermi surface pockets actually exist, the most
likely superconducting symmetry among all possible superconducting states
with an even-frequency order parameter is the $f$ state   $x(x^{2}-3y^{2})
\mathbf{\hat{z}}$.

This is the main result of our paper. However, before concluding, we would like to
remark that an \textit{odd-frequency} triplet $s\mathbf{\hat{z}}$ state, proposed
earlier by us \cite{MDJ+b}, is also compatible with the criteria introduced above. It
has no orbital moment, it is unitary [as opposed to another triplet $s$ state,
$s(\mathbf{\hat{x}\pm }i\mathbf{\hat{y}})],$ it is 2D, and, despite being isotropic, it
has finite DOS at zero energy, giving rise to the observed nonexponential specific heat
and other DOS-sensitive quanitities. As opposed to the $x\mathbf{\hat{z}}$ and
$y\mathbf{\hat{z}}$ case above, $s \mathbf{\hat{x}}$ and $s\mathbf{\hat{y}}$ are also
isotropic and thus do not have any additional disadvantage in terms of the pairing
energy. The ground state in this case is defined either by the spin-orbit induced
magnetic anisotropy (if the easy magnetization axis is in the plane, $s\mathbf{\hat{x }}$
or $s\mathbf{\hat{y}}$ is favored, otherwise $s\mathbf{\hat{z}}$), or by the spin-orbit induced anisotropy of the 
pairing 
interaction.

Finally, we would like comment on the (still controversial) Knight shift
experiments. The absence of a Knight shift decay below $T_{c}$ has been
taken as a decisive argument in favor of the   $(x\pm iy)\mathbf{\hat{z}}$
state in Sr$_{2}$RuO$_{4}$\cite{APM03}. The $f$ state that has emerged from our
discussion also corresponds to Cooper pairs with  spins in the $xy$ plane and
thus to a constant Knight shift. The same is true for the $s\mathbf{\hat{z}}$
state. On the other hand,  $s\mathbf{\hat{x}}$ or $s\mathbf{\hat{y}}$ would
show a reduced, though not exponentially reduced, Knight shift below $T_{c}.$
Of the states with an even frequency gap (\#1-25), the states which, formally,
should not show a Knight shift reduction below $T_c$ along some directions
are \#4, 5, 6, 15, 22 and 23.

In conclusion, we have shown that, based on established experimental evidence and a knowledge of the electronic
structure of Na$_{0.3}$\textit{CoO}$_{2}\cdot y$\textit{H}$_{2}$O, many superconducting states that are allowed by
symmetry can be eliminated from consideration.  Of those with an even-frequency gap, only the $f$ states,
$x(x^{2}-3y^{2})\mathbf{\hat{z}}$, and $y(y^{2}-3x^{2})\mathbf{\hat{z}}$ are fully compatible with what is known about
this compound.  The former has no line nodes along the $e_g'$ hole pockets and is therefore energetically favorable,
given the existence of $e_g'$ states at the Fermi level.  In terms of odd-freqency gap states, s$\mathbf{\hat{x}}$,
s$\mathbf{\hat{y}}$, and s$\mathbf{\hat{z}}$ are all consistent with experimental reports.

\end{document}